\documentclass[aps,prx,twocolumn,superscriptaddress]{revtex4-2}
\usepackage{graphicx}% Include figure files
\usepackage{dcolumn}% Align table columns on decimal point
\usepackage{bm}% bold math
\usepackage[normalem]{ulem}
\usepackage{xcolor}
\usepackage[utf8]{inputenc}
\usepackage[T1]{fontenc}
\usepackage{mathptmx}
\usepackage{etoolbox}
 \usepackage{amsmath}
 \usepackage{booktabs}

\usepackage[colorlinks=true, linkcolor=blue, urlcolor=blue, citecolor=blue]{hyperref}

\makeatletter
\def\@email#1#2{%
 \endgroup
 \patchcmd{\titleblock@produce}
  {\frontmatter@RRAPformat}
  {\frontmatter@RRAPformat{\produce@RRAP{*#1\href{mailto:#2}{#2}}}\frontmatter@RRAPformat}
  {}{}
}%
\makeatother
\begin{document}

\preprint{AIP/123-QED}

\title[Improved two-cell echo diffusing wave spectroscopy]{
Expanding the reach of diffusing wave spectroscopy and tracer bead microrheology}
% Force line breaks with \\
\author{M. Helfer}
\affiliation{Department of Physics, University of Fribourg, Chemin du mus\'e{}e 3, 1700 Fribourg, Switzerland}
\author{C. Zhang}%

\affiliation{Department of Physics, University of Fribourg, Chemin du mus\'e{}e 3, 1700 Fribourg, Switzerland}%
\author{F. Scheffold}
\affiliation{Department of Physics, University of Fribourg, Chemin du mus\'e{}e 3, 1700 Fribourg, Switzerland} \email{Frank.Scheffold@unifr.ch}

\date{\today}% It is always \today, today,

\begin{abstract}
Diffusing Wave Spectroscopy (DWS) is an extension of standard dynamic light scattering (DLS), applied to soft materials that are turbid or opaque. The propagation of light is modeled using light diffusion, characterized by a light diffusion coefficient that depends on the transport mean free path \(\ell^\ast\) of the medium. DWS is highly sensitive to small particle displacements or other local fluctuations in the scattering properties and can probe sub-nanometer displacements. Analyzing the motion of beads in a viscoelastic matrix, known as one-bead microrheology, is one of the most common applications of DWS. Despite significant advancements since its invention in the late 1980s, including two-cell and multispeckle DWS, challenges such as merging single- and multispeckle data and limited accuracy for short correlation times persist. Here, we address these issues by improving the two-cell echo DWS scheme. We propose a calibration-free method to blend and merge echo and two-cell DWS data and demonstrate the use of an exponential basis fit to enhance data quality, in particular at very short times. Building on this, we introduce stable corrections for bead and fluid inertia, significantly improving the quality of microrheology data at high frequencies.

\end{abstract}

\maketitle

\section{Introduction\label{sec:1}}
Diffusing Wave Spectroscopy (DWS) is a standard light scattering technique used to study the dynamics of colloids, colloid-polymer mixtures, surfactant solutions, and other soft matter systems that display interesting complex relaxation dynamics~\cite{weitz1993diffusing,pine1990diffusing,PineWeitz1993,maret1997diffusing,harden2001recent,dasgupta2002microrheology,willenbacher2007broad,galvan2008diffusing,fahimi2017diffusing,xing2018microrheology,lorusso2021recent,dennis2024diffusing,FSinlindner2024neutrons}. Introduced in the late 1980s~\cite{maret1987multiple,PhysRevLett.60.1134} DWS is an extension of the standard dynamic light scattering (DLS) approach~\cite{berne2000dynamic}. In contrast to DLS, it is applied to materials that are turbid or opaque. In such materials, light is not scattered only once, as is the case in DLS, but multiple times. The propagation of light is often modeled using light diffusion characterized by a light diffusion coefficient that depends on the transport mean free path $\ell^\ast$ of the medium. Coherent light becomes dephased due to the motion of the scatterers in the medium~\cite{scheffold2001diffusing,boas1995scattering,boas1997spatially}, which can be quantified by analyzing the intensity time autocorrelation function (ICF) of the scattered light, analogously to the case of DLS.
%%%%%%%%%%%%
\newline \indent DWS is commonly applied to samples such as dense suspensions and emulsions~\cite{qiu1990hydrodynamic,rojas2002diffusing,li2023two,dennis2024diffusing,kim2019diffusing,lorusso2021recent,xu2023jamming}. Alternatively, it can be applied to transparent samples by adding tracer beads that sense the dynamic environment of the medium~\cite{mason1995optical,gardel2005microrheology,willenbacher2007broad,harden2001recent,waigh2005microrheology,furst2017microrheology,zhang2017improved,li2023two,FSinlindner2024neutrons}. The advantage of DWS over DLS is that it can be applied to turbid media that are not accessible to classical DLS, in regimes where multiple scattering suppression techniques such as 3D-cross-correlation fail~\cite{block2010modulated}. Additionally, DWS is more sensitive than DLS to small particle displacements or other local fluctuations in the scattering properties. DWS can probe subnanometer displacements, as demonstrated by Weitz et al. in 1989 shortly after the method was introduced~\cite{weitz1989nondiffusive}. The high sensitivity of DWS is especially useful when tracer beads are embedded in a viscoelastic medium of interest.  DWS can measure viscoelastic solids with elastic moduli of up to 100 kPa~\cite{furst2017microrheology,wyss2001diffusing} and can study relaxations on microsecond timescales, such as Rouse-Zimm and bending modes of polymer dynamics~\cite{gittes1998dynamic,dasgupta2002microrheology,willenbacher2007broad}. 
%%%%%%%%%%%%
\newline \indent Analyzing the motion of beads in a viscoelastic matrix is known as (single) bead microrheology~\cite{harden2001recent,gardel2005microrheology,furst2017microrheology}, which is one of the most common applications of DWS. Pioneering work by Mason, Weitz and others showed that the mean squared displacements of beads $<\Delta \vec{r}^2(t)>$ can be directly converted into the linear viscoelastic spectrum, or complex modulus $G^\ast(\omega)=G^\prime(\omega)+iG^{\prime\prime}(\omega) $, of the materials if the radius of the beads is known~\cite{mason1995optical,gittes1997microscopic,levine2001response}. DWS provides access to this information from submicrosecond time scales to seconds and minutes, opening opportunities in rheology to study viscoelastic moduli over an enormous range of frequencies with a single measurement~\cite{willenbacher2007broad}. Many reviews of the theory underlying the method, along with numerous examples of its applications, have been published. We refer the reader to the literature for further details~\cite{weitz1993diffusing,PineWeitz1993,maret1997diffusing,harden2001recent,waigh2005microrheology,furst2017microrheology,FSinlindner2024neutrons}.
%%%%%%%%%%%%
\newline \indent Since the invention of DWS, several substantial improvements have been made to the method. The original approach relied on time averages of the intensity correlation function (ICF), which requires the sample to be in a liquid state. Two-cell DWS, introduced in 2000~\cite{romer2000sol,scheffold2001diffusing}, and other ensemble averaging approaches~\cite{pusey1989dynamic,xue1992nonergodicity,nisato2000diffusing} provided access to short correlation times for solid nonergodic systems. Multispeckle DWS, using either a camera or the echo scheme, was introduced in the early 2000s and provides access to longer correlation times for highly viscous or viscoelastic solids such as glasses and gels~\cite{knaebel2000aging,cardinaux2002microrheology,viasnoff2002multispeckle,zakharov2006multispeckle,pham2004ensemble,zhang2022echo,alexander2007diffusing,xue2025investigating,skipetrov2010noise}. In 2017, an automated method was developed to determine the mean free paths of optical transport and absorption for absorbing and non-absorbing samples~\cite{zhang2017improved}.
These advancements have made it possible to access the ICF over a range from typically 10ns to tens of seconds or more, while keeping the measurement time on the order of a few minutes. Only when studying even longer relaxation times does the measurement time need to be further increased. 
\newline \indent Despite these significant advances, shortcomings in the implementation of these methods can still lead to errors or reduced data quality. Moreover, access to very short correlation times is hindered by photon shot noise and the need to normalize the ICF, which, in its current implementation, introduces bias.
\newline \indent In the present work, we address these problems and propose two substantial improvements to the implementation of the two-cell echo DWS measurement scheme, which is the most common implementation of DWS. We present a calibration-free method to blend and merge echo and two-cell DWS measurements. Additionally, we demonstrate that using an exponential basis fit directly to the intensity correlation function (ICF) 
can substantially improve the quality of the data. On the basis of this approach, we further demonstrate that a full correction for the bead and fluid inertia can be applied in a stable manner, further enhancing the microrheology data quality in the important high-frequency regime.

We note that a different approach, diffuse correlation spectroscopy using fast single-photon avalanche diode (SPAD) arrays, has made substantial progress over the past decade. The pixel count of such arrays has increased markedly and, although costs remain high, they have decreased significantly. Sie \textit{et al.} \cite{sie2020high,wayne2023massively} demonstrated multispeckle diffuse correlation spectroscopy with a \(64 \times 64\) SPAD array comprising \(4096\) pixels and a temporal resolution of \(4~\mu\mathrm{s}\), while Wayne \textit{et al.} reported a \(500 \times 500\) SPAD camera achieving a full-frame temporal resolution of \(10~\mu\mathrm{s}\). Although not yet widely available and still under development, the SPAD approach is becoming increasingly competitive.
\section{Method}
Conventional photon correlation spectroscopy (PCS), as used in DLS or DWS, is based on the time-averaged photon correlation function recorded by a single-photon detector and processed using a hardware electronic or software correlator ~\cite{berne2000dynamic}. Echo-DWS employs a multispeckle detection scheme to accurately measure the correlation function also in the presence of slow relaxation dynamics, typically for times equal to or longer than tenths of a second. Multispeckle DWS is achieved by illuminating an optical diffuser (e.g., ground glass) with a laser and then spinning or oscillating the glass at the desired frequency $1/\tau_{\text{Echo}}$~\cite{zakharov2006multispeckle,zhang2022echo}. The modulated speckle beam is used to illuminate the sample of interest.
We denote by 
 \begin{itemize}
    \item \( g_2(t) \): the DWS transmission intensity correlation function (ICF) of the sample,
    \item \( g_2^{\mathrm{TC}}(t) \): the ICF measured in a two-cell (TC) configuration, where a slowly moving diffuser ('second cell') is placed in the beam path together with the sample,
    \item \( g_2^{\mathrm{Echo}}(t) \): the ICF obtained from correlation echoes recorded with a rotating or oscillating diffuser with period \(\tau_{\text{Echo}}\),
    \item \( g_2^{\mathrm{2nd}}(t) \): the two-cell ICF for a solid sample,  where \( g_2(t)\equiv 2 \), and a slowly moving diffuser. The decay time  \(\tau_{\text{TC}}\) is controlled by the $\mathrm{2nd}$ cell motion.
\end{itemize}

In a typical experiment, the properly averaged echo correlation function is combined with a two-cell DWS measurement~\cite{romer2000sol,viasnoff2002multispeckle,scheffold2001diffusing}, where the ground glass is slowly displaced to force the decay of the intensity autocorrelation function (ICF) of the sample on a typical time scale $\tau_{TC}$. These approaches ensure proper ensemble averaging, achieved by illuminating the sample with a larger number of different speckle fields. Both measurements are then combined to obtain a single ICF spanning correlation times from typically 12.5 ns to ten seconds or more, covering an impressive nine orders of magnitude in relaxation times. However, the method, as it is currently used, suffers from two main limitations, both of which we address here. 

\subsection{Blending and Merging the Two-Cell and Echo Signals}
The shortest time accessible to Echo-DWS is determined by the periodicity of the rotation or oscillation of the diffuser. Since this requires mechanical oscillation or rotatation of an object weighing a few grams with high precision, the minimum accessible echo time is currently limited to just under a tenth of a second. Realizing somewhat shorter times is possible, but the procedure tends to be less stable and more prone to drift, requires more startup time for stabilization, or requires more complex and expensive equipment to provide the required accuracy. In this study, we used an already elevated frequency of 25 Hz, corresponding to the first echo located at \(\tau_{\text{Echo}} = 0.04\) seconds. Commercial instrumentation, such as the DWS RheoLab (LS Instruments, Switzerland),  typically operates with echo-correlation times that start at $0.1-$ seconds.
%%%%%%%%%%%%
\newline \indent To be able to merge the Echo-DWS and the two-cell DWS intensity correlation functions, $g_2(t)-1$, the forced decay of the latter must be set to a time similar to or greater than \(\tau_{\text{Echo}}\). For comparison, in the original two-cell DWS study by Romer et al.~\cite{romer2000sol}, the value of \(\tau_{\text{TC}}\) was six times greater than the final data point reported for \( g_2(t) - 1 \).
Using the multiplication rule proposed in ref.~\cite{scheffold2001diffusing}, the relation 
\begin{equation}
g_2^{\text{TC}}(t)-1 = \big(g_2(t) - 1\big) \times \big(g_2^{\text{2nd}}(t) - 1\big) \label{twocell}
\end{equation}
can be applied to the two-cell data. By dividing the measured \(g_2^{\text{TC}}(t) - 1\) by \(g_2^{\text{2nd}}(t) - 1\), we recover the true intensity correlation function \(g_2(t) - 1\), where \(g_2^{\text{2nd}}(t) - 1\) is obtained independently through a long calibration measurement on a solid reference sample, such as a slab made of Teflon or colloidal beads embedded in a solid epoxy matrix. We note that for clarity, in this section we ignore the influence of the experimental coherence factor $\beta$ and set it to one.
\newline \indent The maximum time to which the true intensity correlation function can be reliably extracted from TC-DWS depends on the precision of the division based on Eq.~\eqref{twocell}. Unfortunately, the accuracy is often not very high, primarily due to the limited time available to average during the recording of \(g_2^{\text{TC}}(t) - 1\). For example,  for \(\tau_{\text{TC}} = 0.3\), with a total measurement time of 300 seconds this implies that the time average of $g_2^{\mathrm{TC}}(t)-1$ covers $\sim 300/0.3 = 1000$ statistically independent speckle realizations. The statistical noise scales as $1/\sqrt{1000} \simeq 3\%$~\cite{skipetrov2010noise}; see also Table~S1.
Thus, in practice, the forced decay of \(g_2^{\text{TC}}(t) - 1\) exhibits significant variations, and dividing by the measurement of the reference sample, \(g_2^{\text{2nd}}(t) - 1\), frequently results in visible perturbations of the ICF around \(\tau_{\text{Echo}}\). Moreover, if
\(g_2(t) - 1\) is joined with \(g^\text{Echo}_2(t) - 1\) at \(\tau_{\text{Echo}}\) without data overlap, any deviation of the experimental \(g_2(t) - 1\) from the properly ensemble-averaged value leads to a noticeable kink in the final correlation function used for further analysis; see also Fig. S1. If data overlap is to be achieved using the standard approach, one would need to increase \(\tau_{\mathrm{TC}}\), requiring a manifold extension of the measurement time, which is not practical.
%%%%%%%%%%%%
\newline \indent  Here, we propose an alternative approach based on dividing the measured correlation function by the echo correlation function. The latter is, up to a prefactor $A$, proportional to the true correlation function. Furthermore, the echo-ensemble average is generally accurate due to the large number of independent realizations probed during a single echo interval~\cite{zakharov2006multispeckle}. Thus, 
\begin{equation}
\frac{g_2^{\text{TC}}(t) - 1}{g_2^{\text{Echo}}(t) - 1} = A \big(g_2^{\text{2nd}}(t) - 1\big) = A e^{-(t/\tau_{\text{TC}})^p}.\label{eq:echotransdecay}
\end{equation}
The decay caused by the second cell, $g_2^{\text{2nd}}(t) - 1$ , is modeled as an exponential function, \( e^{-(t/\tau_{\text{TC}})^p} \). Due to the unidirectional motion of the diffuser \( p = 2 \) and \( \tau_{\text{TC}} = \text{const.} \), see Fig.~S4.  
However, for finite measurement times, both values, $A$ and $p$, can fluctuate around the mean value. For the examples shown in the main text that study a viscoelastic liquid, we find that \( p = 2 = \text{const.} \) can be used, while for the viscoelastic solid reported in the supplemental material, fluctuations in \( p \) become noticeable, and therefore both \( \tau_{\text{TC}} \) and \( p \) have to be treated as adjustable parameters.
\newline \indent  Using this procedure, we obtain an effective second-cell decay characteristic of the measurement we carried out, removing the need for a calibration sample. We extract \(A\) and \(\tau_{\text{TC}}\) from a fit to the data over, e.g., six data points, and then use \(\tau_{\text{TC}}\) to correct \(g_2^{\text{TC}}(t) - 1\) and \(A\) to adjust the amplitude of \(g_2^{\text{Echo}}(t) - 1\). Finally, we combine the data over the fitting region and then append the remaining data points from the echo measurement. This approach ensures a smooth crossover between the two measurements without discontinuities.  
%By interpolation, we map the data for $t\ge \tau_{\text{Echo}}$  onto the original multiple-tau channel layout of the time-averaged ICF, preserving the logarithmic channel distribution on the time axis across the entire dataset.
\subsection{Analysis of the short-time correlation function decay}
One of the key features of DWS is the ability to measure tiny displacements of particles or other local rearrangements. DWS enables measurements of sub-diffusive motion on nanometer length scales using visible light as a probe. In DWS-based microrheology, relaxations on (sub)microsecond time scales correspond to frequencies in the viscoelastic spectra of \(10^6 \, \text{rad/s}\) and above, which no other laboratory technique can measure. For comparison, piezo-driven squeeze-flow devices can reach frequencies up to approximately \(10^4\,\text{rad/s}\)~\cite{schroyen2020bulk}. Torsional resonance oscillators extend the accessible frequency range into the \(10^5\,\text{rad/s}\) regime \cite{fritz2003characterizing};  however, rheological measurements at such high frequencies are often limited by experimental constraints and are therefore impractical in most routine applications. Displacements of particles on these time scales are sub-Brownian, which means they are increasingly affected by the finite inertia of the tracer particles or beads and the surrounding fluid, which we will discuss in the next section. First, we focus on the accurate measurement of the short-time decay. DWS provides \( \left[k \left(L/\ell^\ast\right)\right]^2 \left<\Delta \vec{r}^2(t)\right> \)
over a specific range determined by the precision of the measurement, where \( k = 2\pi n/\lambda \) is the wavenumber of scattered light in the host medium with a refractive index \( n \).
By increasing the optical density \( L / \ell^\ast \) of the sample - either by increasing the cell thickness $L$ or by lowering \( \ell^* \) through an increase in the concentration of tracer beads - one can shift the window of accessible mean squared displacements to smaller values (and shorter time scales). However, this comes at the expense of reduced sensitivity to larger displacements.
\newline \indent For \(t \to 0\) the ICF tends to a constant value of \(g_2(0) - 1 = \beta\) where \(\beta \lesssim 1\) denotes the coherence factor of the detection fiber optics and $\beta + 1$ corresponds to the zero-time \emph{intercept} of the ICF.
In a typical DWS microrheology experiment, for short relaxation times up to some tens or hundreds of microseconds, the bead motion often leads to relatively minor decays of the measured ICF, often only by a few per cent. Extracting $\left<\Delta \vec{r}^2(t)\right>$ from these minor decays is strongly affected by the accuracy with which one can determine \(\beta \) and by the significant photon shot noise at short correlation lag times~\cite{schatzel1983noise}. Normalization to determine $\beta$ is usually performed by extrapolating the measured \( g_2(t)\) to \( t = 0 \) using a linear or polynomial fitting function. Once \(\beta\) is known, the normalized ICF is used to determine, e.g. \(\left<\Delta \vec{r}^2(t)\right>\). 
This standard procedure suffers from two problems. First, the choice of the polynomial order used for the fit, as well as the time interval used for the fit of the data, influences the exact value of \(\beta\) obtained. Here, it is important to note that, due to the inertia of the beads and the fluid, the short-time dynamics is rather complex even for simple colloidal suspensions in water, and thus no simple temporal scaling of the mean squared displacement $\left<\Delta \vec{r}^2(t)\right>$ can be assumed~\cite{weitz1989nondiffusive,qiu1990hydrodynamic,yoon2018brownian}. Second, since the raw ICF is normalized by \(\beta\) but is used as is for further analysis, the experimental shot noise is not removed. As a result, the data fluctuate around the mean and reach values greater than one, which is unphysical and leads to invalid data points for the bead mean squared displacement \(\left<\Delta \vec{r}^2(t)\right>\), which introduces further bias. This, in turn, creates difficulties in extracting the rheological information and performing the inertia corrections discussed in the following section.
\subsubsection{Intercept-Adjusted Exponential Basis Fit}\label{sec:expbasfit}

To obtain a smooth representation of the experimental data we approximate the electric field autocorrelation function \( g_1(t) \) using a fit with fixed, logarithmically spaced decay times:
\begin{equation}\label{Eq:ExBFitg1}
g_1(t) = \sum_i P_i\, e^{-t / \alpha_i},
\end{equation}
where the basis functions are exponentials with fixed decay times \(\alpha_i\), and the amplitudes \(P_i \ge 0\) are determined by fitting. 
We employ \( N \) logarithmically spaced decay times defined as
\begin{equation}
\alpha_i = 10^{-7}\,\mathrm{s} \, \times e^{\frac{i-1}{u}},
\qquad i = 1, \dots, N,
\end{equation}
$\alpha_N$ is chosen to be one or two orders of magnitude larger than the time window of interest. Solving for \(u\) gives
$u = (N-1)/\ln\left(\alpha_{N}/\alpha_0\right).$ For viscoelastic solid samples exhibiting a stable plateau, the fit can be further improved by including a constant offset term in Eq.~\eqref{Eq:ExBFitg1}.

Here we use $N=45$ and $u=2$ $\left(\alpha_N=360s\right)$ to ensure robustness, but tests with smaller values yield similar results at a significantly lower computational cost as long as $N\ge10$, see Fig.~\ref{fig:ContinFit}~(c). This exponential basis approach provides a stable and flexible reconstruction of \(g_1(t)\), enabling an accurate representation even at very short times. Applying the Siegert relation, the intensity correlation function is modeled as:
\begin{equation}
g_2(t) - 1 = \left( \sum_i P_i e^{-t / \alpha_i} \right)^2
\end{equation}
This formulation implicitly adjust the intercept via
\begin{equation}
\beta+1=g_2(0) = 1+\left( \sum_i P_i \right)^2
\end{equation} and thus allows us to normalize the ICF, avoiding the need for separate normalization or intercept fitting. The fitting procedure minimizes the mean squared deviation between the modeled and measured ICF:
\begin{equation}\label{eq:global}
\text{Error} = \frac{1}{{N^\prime}} \sum_{j=1}^{N^\prime} \left[ \left( \sum_i P_i e^{-t_j / \alpha_i} \right)^2 - \left( g_2(t_j) - 1 \right) \right]^2
\end{equation}
where $j=1 \to N^\prime$ is the data-point index in the multiple-tau channel layout of the ICF, and $N^\prime$ is the total number of data
points.

After obtaining the optimal weights \( P_i \), the model is used to reconstruct a smoothed, intercept-corrected ICF:
\begin{equation}
g_2^{\text{fit}}(t) - 1 = \left( \sum_i P_i e^{-t / \alpha_i} \right)^2/\left( \sum_i P_i \right)^2
\end{equation}

Finally, the MSD is extracted by numerically inverting this fitted correlation function using the full Diffusing Wave Spectroscopy (DWS) model in transmission geometry, as given in Eq.~\eqref{eq:msd}. 
\newline \indent We emphasize that this fitting procedure does not constitute an additional smoothing step in the DWS microrheology analysis; rather, it replaces the conventional smoothing fit applied to the mean-square displacements \cite{mason1995optical}, offering the advantages discussed above.

\subsection{One-bead microrheology with inertia corrections}
The time relaxation of the intensity correlation function (ICF) is determined by the motion of the beads and the light scattering process. Precise analytical formulas have been developed to extract the bead mean square displacement (MSD) from the measured DWS intensity autocorrelation functions in transmission geometry.  The measured mean square displacement, in turn, is determined by the linear viscoelastic properties of the host medium and, at short times, also by the inertia of the bead and the fluid. Particle interactions, convection or flow, and phase separation can also influence the particle motion, but these effects will not be discussed here~\cite{PineWeitz1993,FSinlindner2024neutrons}. 
Ignoring such effects, the analysis procedure proceeds as follows: First, we extract the  mean squared displacement (MSD), $\langle \Delta\vec{r}^2(t) \rangle$, from the ICF.    Next, we remove the contributions due to bead and fluid inertia to obtain the unperturbed viscoelastic spectra of the (complex fluid) host medium.
The different steps and approximations made will be discussed below. In addition, we will highlight the importance of clean and low-noise data to stably perform this analysis.

\subsubsection{Mean squared displacement}
We first convert the experimental data to the MSD using Eq.~\eqref{eq:msd} taken from ref.~\cite{PineWeitz1993}, assuming that the slab thickness \( L \), the sample's transport mean free path \( \ell^\ast \), and the wavenumber $k$ are known. Since we study non-absorbing samples, we neglect the effects of absorption, but these can be easily incorporated if required~\cite{zhang2017improved}.
\begin{widetext}
\begin{equation}\label{eq:msd}g_2(t)-1=\left\{ \frac{  \frac{L/\ell^\ast+4/3}{z_0/\ell^\ast+2/3}\left(\sinh\left[\frac{z_0}{\ell^\ast}\sqrt{k^2 \langle \Delta\vec{r}^2(t) \rangle}\right] + \frac{2}{3} \sqrt{k^2 \langle \Delta\vec{r}^2(t) \rangle} \cosh\left[\frac{z_0}{\ell^\ast}\sqrt{k^2 \langle \Delta\vec{r}^2(t) \rangle}\right]\right)}{\left(1 + \frac{4}{9} k^2 \langle \Delta\vec{r}^2(t) \rangle \right) \sinh\left[\frac{L}{\ell^*} \sqrt{k^2 \langle \Delta\vec{r}^2(t) \rangle}\right] + \frac{4}{3} \sqrt{k^2 \langle \Delta\vec{r}^2(t) \rangle} \cosh\left[\frac{L}{\ell^*} \sqrt{k^2 \langle \Delta\vec{r}^2(t) \rangle}\right]}\right\}^2\end{equation}
\end{widetext}
If internal reflections are absent or neglected, $z_0 / \ell^\ast \simeq 2/3$. In our case, we consider a glass cuvette in air; thus, $z_0 / \ell^\ast \simeq 2$, a value derived from the work of Durian et al.~\cite{lemieux1998diffusing}. The larger $L/ \ell^\ast$, the less important the choice of $z_0$ becomes~\cite{FSinlindner2024neutrons}. By numerical inversion of Eq.~\eqref{eq:msd} for each value of \( t \), we obtain a precise value of \( \left<\Delta \vec{r}^2(t)\right> \) for each value of $g_2^{\text{fit}}(t) - 1$, with no fitting or ambiguity involved in the process.

\subsubsection{Microrheology Routine}\label{sec:MicrorheologyRoutine}
We use the following microrheology routine, based on Mason's work~\cite{mason2000estimating}, to determine the apparent complex modulus (inertia affected) \( Z^\ast(\omega) \) from the mean squared displacement data \( \langle \Delta r^2(t) \rangle \), determined in the previous step, as follows:

\begin{equation}\label{eq:MasonMR}
\alpha(\omega) = \left.\frac{\rm{d} \ln \langle \Delta \vec{r}^2(t) \rangle}{\rm{d} \ln t}\right|_{t = 1/\omega},
\end{equation}

\begin{equation}\label{msdapproax}
i \omega F_\omega\{\langle \Delta \vec{r}^2(t) \rangle\} \approx \langle \Delta \vec{r}^2\left(\frac{1}{\omega}\right) \rangle \, \Gamma[1 + \alpha(\omega)] i^{-\alpha(\omega)},
\end{equation}

where $t$ is substituted by $1/\omega$, and finally 

\begin{equation}\label{eq:appcomplex}
Z^\ast(\omega) = \frac{k_B T}{\pi a \omega F_\omega\{\langle \Delta \vec{r}^2(t) \rangle\}}.
\end{equation}
where $a$ denotes the radius of the tracer beads. Eq.~\eqref{eq:appcomplex} is also known as the generalized Stokes-Einstein equation (GSE) (in the Fourier domain)~\cite{mason2000estimating,furst2017microrheology}.

We note that Evans et al. developed a direct conversion algorithm from the mean-squared displacement (MSD)—or equivalently, from the creep compliance (see, e.g., Eq. (3.99) in~\cite{furst2017microrheology})—to the storage and loss moduli~\cite{evans2009direct,tassieri2012microrheology}. This method circumvents the approximation used in Eq.~\eqref{msdapproax}. It was later refined by Tassieri et al.~\cite{tassieri2012microrheology}. Figure S4 shows that our approach and the direct conversion yield nearly identical results. The method by Evans et al. tends to produce noisier data, while the refined version by Tassieri et al. mitigates this issue by avoiding an approximation that can introduce artificial scatter~\cite{tassieri2018comment}. Small residual differences remain, but a detailed analysis of these discrepancies lies beyond the scope of the present work and should be addressed in future studies.

\subsubsection{Inertia correction}
At very short times, the common description of thermal Brownian motion breaks down, a phenomenon that has been well known since Einstein's pioneering work on the subject in 1905~\cite{einstein1905motion}. The transition from ballistic to Brownian dynamics, which is strongly affected by inertia and hydrodynamics, has attracted renewed attention more recently~\cite{hinch1975application,huang2011direct,madrid2023ballistic,coglitore2017transition}. 
Neglecting these contributions in the microheology analysis can result in significant errors. Properly accounting for these effects at short times, corresponding to high frequencies (\(\omega\)) when determining the complex modulus \(G^\ast(\omega)\), enables the extraction of critical data that would otherwise be challenging or impossible to obtain.
\newline \indent The inertia correction formula used in the present study, Eq.~\eqref{eq:inertiacorr}, is based on the work of Dom\'{\i}nguez-Garc\'{\i}a et al.~\cite{dominguez2014accounting}. The complete theoretical treatment applied here was first developed and published by Schieber et al.~\cite{indei2012competing}. Prior to that, an (iterative) approximative algorithm to correct for inertia effects was developed by one of us in Willenbacher et al.~\cite{willenbacher2007broad} and similarly employed by Mizuno et al.~\cite{mizuno2008active}.
From the apparent complex modulus \( Z^\ast(\omega) \), extracted from the experimental data, we calculate the complex modulus of the host medium as follows:

\begin{widetext}
\begin{equation}\label{eq:inertiacorr}
G^\ast(\omega) = Z^\ast(\omega) + \frac{m^\ast \omega^2}{6 \pi a} + \frac{a^2 \omega^2}{2}
\left[ \sqrt{\rho^2 - \frac{2 \rho}{3 \pi a^3} \left(\frac{6\pi a}{\omega^2}Z^\ast(\omega) + m^\ast\right)} - \rho \right],
\end{equation}
\end{widetext}
The effective mass of the tracer beads is:

\begin{equation}
m^\ast = \frac{4 \pi a^3}{3} \rho_b + \frac{2 \pi a^3}{3} \rho,
\end{equation}
where \( \rho_b \) is the bead density and \( \rho \) is the medium density in \(\text{kg/m}^3\). The resulting \( G^\ast(\omega) \) is a complex functions, which can be evaluated using computational tools such as Mathematica (Wolfram, USA)~\cite{mathematica} or Python (Python Software Foundation)~\cite{python}. 
We can then obtain the following:
\begin{itemize}
    \item The absolute value \( |G^\ast(\omega)| \),
    \item The storage modulus \( G^\prime(\omega) = \text{Re} \, G^\ast(\omega) \),
    \item The loss modulus \( G^{\prime\prime}(\omega) = \text{Im} \, G^\ast(\omega) \).
\end{itemize}
\section{Results - Experimental validation} 
\subsection{Sample}
We study DWS microrheology on a worm-like micellar sample discussed extensively in previous work~\cite{fischer1997rheological,cardinaux2002microrheology,willenbacher2007broad,oelschlaeger2009linear}. To do this, we prepare an aqueous solution of cetylpyridinium chloride and sodium salicylate (100 mM CPyCl - 60 mM NaSal). We mix the sample with polystyrene beads with a diameter of 420 nm to a final concentration slightly below 1\% in volume.  We determine the transport mean free path $\ell^\ast$ using the commercial 'DWS RheoLab' instrument (LS Instruments, Switzerland)~\cite{zhang2017improved}, and, using Mie scattering theory, correct the obtained value for the slight mismatch in laser wavelength. We prepare the sample such that the micellar concentration relative to the aqueous solvent phase is kept constant. The sample displays nearly perfect Maxwellian behavior at low frequencies or slow relaxation times and exhibits the characteristic dynamics of polymer relaxations at short times. It is a strongly viscoelastic liquid with a terminal relaxation time that reacts to temperature over a wide range. Here, we use it as a model system to demonstrate our improved implementation of two-cell echo diffusing wave spectroscopy for different terminal relaxation times larger than or comparable to $\tau^{\mathrm{TC}}$. 

\subsection{Experimental Setup}
We carry out DWS experiments in transmission geometry, where an expanded incident laser beam illuminates the sample contained in a rectangular glass cuvette, with a width larger than its thickness $L$. The diffusely scattered light is recorded on the opposite side, in transmission, using a single-mode optical fiber. We use a classical custom-made DWS setup with a Helium-Neon gas laser operating at a 632.8~nm wavelength emission. The collimated laser beam is directed towards a 12.5mm diameter holographic diffusor (Edmund Optics, UK) with a 5$^\circ$ diffusing angle.
Compared to ground- or opal-glass diffusers, holographic diffusers offer higher light transmission and a more even distribution of light. They provide control over the diffusion angle, eliminating the need for an additional collimation lens between the diffuser and the sample. In our setup the diffusor is mounted on a Voice Coil Actuated Flexure Scanner (VCFL35, Thorlabs, USA). The scanner is operated at 25 Hz using a function generator. The light diffracted from the diffuor in the forward direction forms a random speckle field that we use to illuminate the sample cuvette.  The sample is contained in a 12.5 mm wide (10 mm inner width) glass cuvette (Hellma, Germany) with a path length of $L=2$~mm. For the distance chosen for the ground glass from the cuvette (a few centimeters), the speckle beam size is comparable in width to that of the cuvette. We place a black circular aperture with an 8~mm opening in front of the cuvette to avoid illuminating the edges of the cuvette. 
%%% FIGURE %%%%%%%%%%%%%%%%%%%%%%%%%%%%%%%%%%%
%%%%%%%%%%%%%%%%%%%%%%%%%%%%%%%%%%%%%%%%
\begin{figure}\centering
\includegraphics[width=0.9\linewidth]{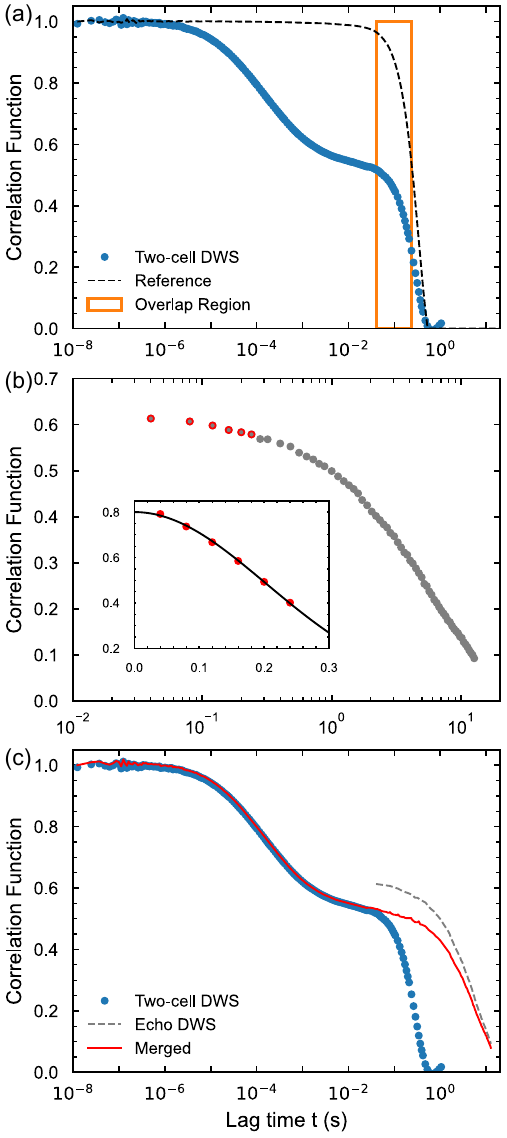}
\caption{\label{fig:EchoTrans} (a) Two-cell correlation function \( g_2^{\mathrm{TC}}(t) - 1 \) recorded over 300 seconds for a wormlike micellar solution at T$=20^\circ$C with polystyrene tracer beads embedded. The overlap region is marked with an orange line. The dashed line shows the correlation function measured for a solid piece of Teflon, recorded over 3 hours and following Eq.~\eqref{eq:echotransdecay} with $A = 0.988$ and $\tau_{\text{TC}} = 0.298$~s. (b) Echo DWS for the same sample recorded over 60 seconds, with data up to 12 seconds. The oscillation period is \( \tau_{\mathrm{Echo}} = 0.04 \,\mathrm{s} \). Inset: Ratio of the two-cell correlation function to the echo correlation function. Line: Best fit to the data with \( A = 0.800 \) and \( \tau_{\mathrm{TC}} = 0.287 \,\mathrm{s} \) using Eq.~\eqref{eq:echotransdecay}. (c) Blending and merging two-cell DWS and echo data. Blue open square: Two-cell correlation function. Dashed line: Echo correlation function. Red line: Merged correlation function obtained by multiplying the echo data by the parameter \( A \) and dividing the two-cell data by the exponential function, Eq.~\eqref{eq:echotransdecay} (up to $t=6 \times \tau_{Echo}$) then merging the two data sets. }
\end{figure}
%%%%%%%%%%%%%%%%%%%%%%%%%%%%%%%%%%%%%%%%%%%%%%%%%%%%%%%
The light transmitted through the multiple scattering sample is detected in the far field on the other side using a single-mode fiber (OZ Optics, Canada). The single-mode fiber is split 50/50 and each arm is connected to an APD single-photon detector (Excelitas, USA). Using a digital correlator (LS Instruments, Switzerland), we calculate the pseudo-cross-correlation function of the scattered photons, which is equal to the photon autocorrelation function, but eliminates detector noise (also known as afterpulsing effects). The setup is similar to the one shown in Figure~1 of reference~\cite{zakharov2006multispeckle}. 
\newline \indent We operate the voice coil scanner in two modes. In slow mode, ground glass motion leads to a slow decay of the intensity correlation function (ICF) on a timescale of $\tau^{\mathrm{TC}}$ but does not affect the short-time dynamics except for providing adequate ensemble averaging of the correlation function, Fig.~\ref{fig:EchoTrans}~(a). We record the intensity correlation function (ICF) using the standard multiple tau $N^\prime$ channel layout (LS Correlator, LS Instruments, Switzerland). We perform an approximate normalization of \( g_2^{\mathrm{TC}}(t) - 1 \) by dividing the data by the values recorded around \( t = 10^{-7} \,\mathrm{s} \), which typically range between 0.9 and 0.95.
  In fast mode, the periodic oscillation of the ground glass with a period of $\tau^{\mathrm{echo}} = 0.04$ sec leads to correlation echoes~\cite{pham2004ensemble,zakharov2006multispeckle,zhang2022echo}, which we detect using the linear correlator scheme configured in the digital correlator. This scheme provides a time resolution of 1 $\mu$s and spans 16 channels centered around $\tau^{\mathrm{echo}} = 0.04$ seconds and multiples thereof. From a Lorentzian fit of the detected correlation echoes, we determined the total echo area—shown in Fig.~\ref{fig:EchoTrans}b rescaled by an arbitrary constant - which gives a robust measure of the intensity correlation function, up to an undetermined prefactor~\cite{pham2004ensemble,zakharov2006multispeckle}. The measurement time for the slow mode were $300~\mathrm{s}$. and $60~\mathrm{s}$ for the fast (echo) mode.
\subsection{Merging two-cell DWS and Echo data}
We first generate an interpolation function for \( g_2^{\mathrm{TC}}(t) - 1 \) with SciPy based on the data recorded using a multiple tau layout of correlation lag times~\cite{schaetzel1991noise}. SciPy employs piecewise polynomial interpolation (cubic interpolation for smoothness) to construct the interpolating function~\cite{scipy}. This function is then used to calculate the values of \( g_2^{\mathrm{TC}}(t) - 1 \) at \( t = \tau^{\mathrm{echo}} \) and its multiples in the linear lag time scheme of the echo correlation function. 
To merge the two datasets, we use the first six echo data points from \( t = \tau^{\mathrm{echo}} = 0.04 \) s to \( t = 6\tau^{\mathrm{echo}} = 0.24 \) s. The choice of six is ad hoc, but it is evident that at least a few overlapping data points are necessary to blend the two measurements effectively, while selecting an excessively large number of overlapping points provides little benefit. For these six points $i=1,2,..,6$ we calculate the ratio $\left[g_2^{\text{TC}}(i \times \tau_\text{Echo}) - 1\right]/\left[g_2^{\text{Echo}}(i \times \tau_\text{Echo}) - 1\right]$. As shown in the inset of Fig.~\ref{fig:EchoTrans}~(b), we find that for \( \tau^{\mathrm{Echo}} \simeq 0.04\)s and our choice of \( \tau^{\mathrm{TC}} \simeq 0.29\)s, this range of data points is well described by Eq.~\eqref{eq:echotransdecay}, and we extract the parameters \( A \) and \( \tau_{\mathrm{TC}} \) from a fit to the data. 
Finally, we multiply the full dataset for \( g_2^{\mathrm{Echo}}(t) - 1 \) by \( A \), divide the dataset \( g_2^{\mathrm{TC}}(t) - 1 \) by \( e^{-(t/\tau_{\mathrm{TC}})^2} \), and truncate it at \( t = 6\tau^{\mathrm{echo}} = 0.24 \,\mathrm{s} \). We merge both data sets into one, and the result is shown as a red line in Figure~\ref{fig:EchoTrans}(c). Our results suggest that as long as \(\tau_{\mathrm{TC}} \geq 6 \tau_{\mathrm{Echo}}\), corresponding to a second-cell ICF decay of no more than \(1/e\) for the last overlapped data point, the parameters \(A\) and \(\tau_{\mathrm{TC}}\) can be extracted accurately and the data can be seamlessly blended and merged. We note that the blending and merging approach introduced here could equally be applied when combining two-cell DWS with camera-based DWS to cover the slow relaxation time window, as discussed in Refs.~\cite{knaebel2000aging,cardinaux2002microrheology}.

 \subsection{DWS microrheology and inertia correction}
\subsubsection{Data preparation}
In the previous section, we demonstrated how to blend and merge two-cell and echo measurements without the need for a calibration sample, achieving a smooth and steady transition between the two datasets. We note that the first few echo data points are relatively widely spaced on the time axis, but they are now combined with the two-cell data over the same time range and instead form a fairly dense data set. Echo data at larger times are densely spaced compared to the multiple-tau channel layout. This outcome is important, as the mean square displacement (MSD) of the particles is typically represented in logarithmic format, and microrheology routines often rely on analyzing derivatives of the logarithm of the mean squared displacement, Eq.~\eqref{eq:MasonMR}. An unevenly spaced distribution of time channels around $\tau_\text{Echo}$ can introduce errors and inconsistencies in data processing.
\subsubsection{Fitting and smoothing}
\begin{figure}\centering
\includegraphics[width=0.9\linewidth]{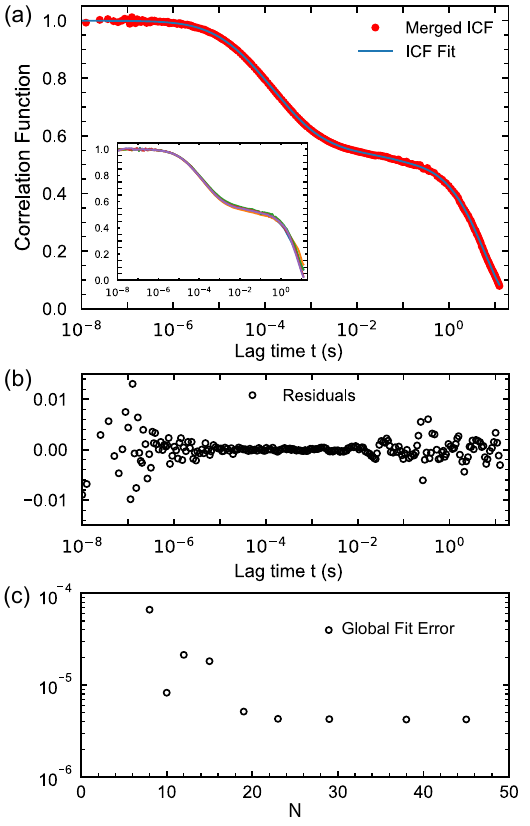}% Here is how to import EPS art
\caption{\label{fig:ContinFit} (a) Fit to the intensity correlation function (ICF). Red circles: Merged correlation function \( g_2(t) - 1 \) as shown in Figure~\ref{fig:EchoTrans}(c). Blue line: Exponential basis fit $g_2^{\text{fit}}(t)$ to the correlation function. The experimental data is normalized using the ICF fit extrapolated to \( t = 0 \). Inset: Five repetitions of the merged correlation function \( g_2(t) - 1 \), normalized using the exponential basis fit. (b) Open circles: Fit residuals $r(t) = g_2(t) - g_2^{\text{fit}}(t)$. (c) Influence of the number
of exponentials on the fit quality. Open circles: global
fit error, Eq.~\eqref{eq:global}, evaluated for different numbers of exponential
terms $N$. }

\end{figure}

We use the merged data set as input for the exponential basis fit, Section~\ref{sec:expbasfit}. Our primary objective is to ach\textbf{}ieve a minimally biased extrapolation to \(t = 0\) and smooth data, enabling the application of the inertia correction and the microrheology algorithm. Using the SciPy interpolation function, we remap the data onto the original multiple-tau channel layout of the digital correlator to prevent oversampling and undue weighting of the longer-time data.
 The merged data set and the fit $g_2^{\text{fit}}(t) - 1$ are shown in Figure~\ref{fig:ContinFit}~(a) together with the residuals, Figure~\ref{fig:ContinFit}~(b). In the inset of Figure~\ref{fig:ContinFit}~a), we show five repetitions of the same experiment to demonstrate its stability and reproducibility, see also supplementary Table 1. We note that without this step, applying the inertia correction using Eq.~\eqref{eq:inertiacorr} would not be feasible without either discarding the high-frequency data of interest or manually adjusting the normalization and smoothing the short-time data, as done in earlier work~\cite{willenbacher2007broad,dominguez2014accounting}.

 \subsubsection{Microrheology}
\begin{figure}\centering
\includegraphics[width=0.9\linewidth]{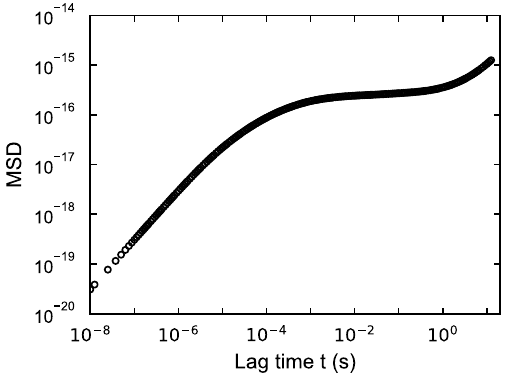} 
\caption{\label{fig:msd}Open black circles: Mean squared particle displacement (MSD) calculated from the intensity correlation function via Eq.~\eqref{eq:msd}.
 }
\end{figure}

\begin{figure}\centering
\includegraphics[width=0.9\linewidth]{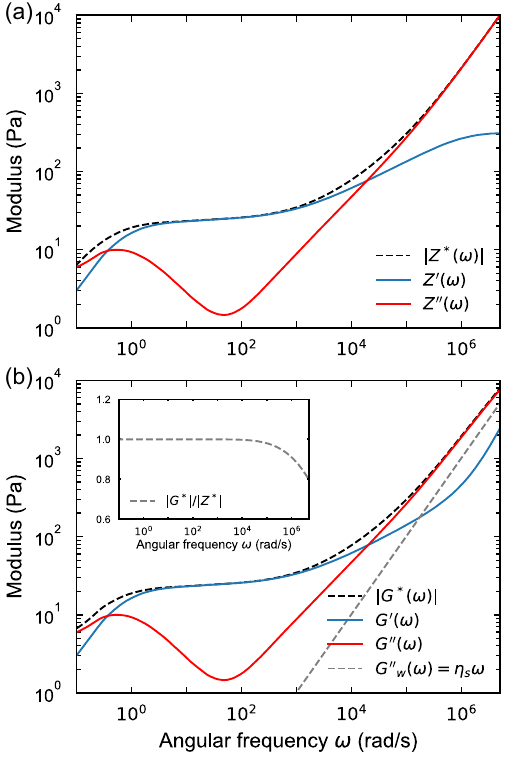}% Here is how to import EPS art
\caption{\label{fig:Moduli} (a) Apparent complex moduli of the micellar solution $Z^\ast(\omega)$ at T$=20^\circ$C, as well as the real and the imaginary part $Z^\prime(\omega),Z^{\prime\prime}(\omega)$, obtained using the Generalized Stokes-Einstein Relation (GSE), Eq.~\eqref{eq:appcomplex}, and Mason's routine to convert the mean squared displacement~\cite{mason2000estimating}. 
(b) Corrected complex modulus of the sample after accounting for bead and fluid inertia, which primarily affects data at frequencies \( \omega \geq 10^{5} \,\mathrm{rad/s} \). The solvent loss modulus $G^{\prime\prime}\left(\omega\right)=\eta_s \omega$ is shown as a dashed line. The data is truncated at \( \omega = 5 \times 10^{6} \,\mathrm{rad/s} \). The inset in panel (b) shows the ratio of uncorrected and corrected modulus $\left|G^\ast\right|/\left|Z^\ast\right|$.
 }
\end{figure}

 We determine the transport mean free path of the samples to \( \ell^\ast \simeq 330 \mu\)m which, together with the known values of \( L = 2 \,\mathrm{mm} \) and \( k = \frac{2\pi n}{\lambda} = 13.2 \,\mu\mathrm{m}^{-1} \), for \( n_{\mathrm{water}} = 1.33 \) is needed to extract the particle mean squared displacement \( \langle \Delta r^2(t) \rangle \) using Eq.~\eqref{eq:msd} and the data is shown in Fig.~\ref{fig:msd}. 
With knowledge of the radius of the beads (\( a = 210 \,\mathrm{nm} \)), we determine the apparent complex modulus \( Z^\ast(\omega) \) using Eq.~\eqref{eq:appcomplex} and the result is shown in Fig.~\ref{fig:Moduli}~(a). Finally, we apply Eq.~\eqref{eq:inertiacorr} to account for the fluid and bead inertia, obtaining the true moduli of the sample, shown in Fig.~\ref{fig:Moduli}~(b), see also Figure S2. The following values are used: \( \rho_b = 1050 \,\mathrm{kg/m^3} \)  (polystyrene bead mass density) and \( \rho \simeq 1000 \,\mathrm{kg/m^3} \) (medium density) and temperature-dependent water viscosity $\eta_s$~\cite{rumble2025crc106}.
\newline \indent In the high-frequency limit, we truncated the data at \( \omega \simeq 5 \times 10^6 \,\mathrm{rad/s} \), above which the ICF decays by less 0.5\% from 1 to 0.995. Using a larger \( L/\ell^\ast \) leads to a faster decay and may provide access to even higher frequencies. However, as seen in Fig.~\ref{fig:Moduli}~(b), the complex modulus gradually approaches the modulus of the solvent \(G^{\prime\prime}\left(\omega\right)= \eta_s \omega \). Thus, the upper bound for a stable DWS high-frequency analysis remains to be determined and is likely to be system dependent. Using the improved DWS implementation, we are already pushing the boundaries of the technique. Still, both the blending and merging procedures and the inertia correction are stable, and the process is highly reproducible.
\newline \indent For a practical test, we apply our method to study the viscoelastic moduli by performing a temperature ramp. Analogous experiments were reported earlier by Oelschlaeger et al., and our data compares quantitatively to this literature data~\cite{oelschlaeger2009linear}.  While the previously published data was truncated at \( \omega \simeq 10^5 \,\mathrm{rad/s} \), here we extend the frequency range reproducibly for all temperatures up to \( \omega \simeq 5 \times 10^6 \,\mathrm{rad/s} \). 
\begin{figure}\centering
\includegraphics[width=.9\linewidth]{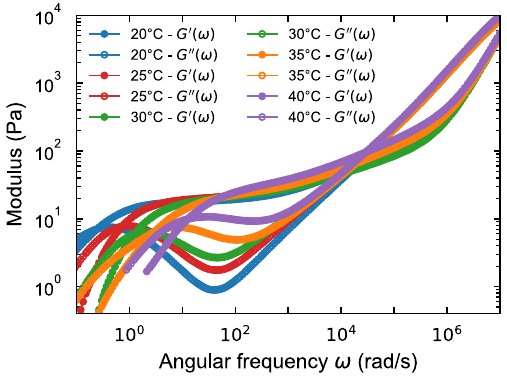}% Here is how to import EPS art
\caption{\label{fig:GModulusTPlot}Microrheology of the micellar solution at different temperatures, ranging from \( T = 20^\circ \mathrm{C} \) to \( T = 40^\circ \mathrm{C} \), illustrating the improved two-cell echo diffusing wave spectroscopy and tracer bead microrheology. }
\end{figure}
Finally, we also demonstrated the method on a viscoelastic solid sample (Figure S5 and S6). To this end, we studied a dense microgel suspension without added tracer beads. The system is identical to the one described in ~\cite{bergman2025free}.

\section{Conclusions} 
In the present work, we presented a new implementation of two-cell echo diffusing wave spectroscopy (DWS) aimed at significantly improving the data quality and stability and thus extending the reach of DWS and DWS-based tracer bead microrheology. We replaced the standard calibration method for two-cell DWS, introduced by Romer et al. in 2000 \cite{romer2000sol}, later used by many others and implemented in commercial DWS equipment~\cite{viasnoff2002multispeckle,zhang2017improved,FSinlindner2024neutrons}, with a calibration-free approach. The new method generically corrects small fluctuations in the terminal decay time arising from imperfect temporal averaging, allowing for a seamless transition between two-cell DWS and echo DWS. This smooth transition, in turn, facilitates a stable conversion of the bead mean square displacement to the complex viscoelastic modulus via the generalized Stokes-Einstein relation (GSER). Combined with the normalization of the correlation function using the exponential basis fit and the elimination of shot noise, the quality of DWS data is substantially improved. This enhancement enables the routine application of published formulas to correct the effects of fluid and tracer bead inertia.
As a result, the method provides unprecedented access to the high-frequency regime~\cite{willenbacher2007broad}. This capability will be beneficial for addressing questions related to the short-time dynamics of polymeric and other soft-matter systems. Furthermore, this approach brings DWS closer to other advanced techniques, such as neutron spin echo spectroscopy, connecting the high-frequency viscoelastic relaxations to sub-microsecond scale dynamics in systems like microgels~\cite{hertle2014internal} or protein hydrogels~\cite{rao2023hindered}.

\begin{acknowledgments} We thank Erika Eiser (NTU Trondheim, Norway) and Andrea Vaccaro (LS Instruments, Switzerland) for valuable discussions. This work was financially supported by the Swiss National Science Foundation through project grant \#10000141 and \#10002524. We thank the referees for their constructive feedback, and in particular the second referee for helpful suggestions on improving the fitting of the ICF.
\end{acknowledgments}

\bigskip 
\section*{Author Contributions} F.S. designed the study and wrote the original Mathematica code. M.H. and C.Z. conducted the experiments. C.Z. designed the exponential basis fit. M.H. implemented and tested the Python package (pyDWS). All authors participated in the data analysis. F.S. wrote the manuscript with contributions from all authors.

\bigskip
\section*{Data Availability Statement}
All experimental data and model output discussed in the manuscript have been uploaded to the repository Zenodo (\href{https://doi.org/10.5281/zenodo.17426137}{17426137}). The source code for the pyDWS Python package can be found on \href{https://github.com/ManuelH26/pyDWS}{GitHub/pyDWS} and can also be installed directly from \href{https://pypi.org/project/pyDWS}{PyPi/pyDWS}. All additional data sets generated and / or analyzed during the current study are available from the corresponding author upon reasonable request.

\end{document}